\documentclass[english,prl, twocolumn, twosided]{revtex4}

\usepackage[latin1]{inputenc}
\usepackage{graphicx}

\makeatletter

\usepackage{babel}
\makeatother
\begin{document}

\title{Measuring the size of a Schr\"odinger cat state}

\author{Florian Marquardt, Benjamin Abel, and Jan von Delft}

\affiliation{Physics Department, Arnold Sommerfeld Center for Theoretical Physics,
and Center for NanoScience, Ludwig-Maximilians Universit\"at, Theresienstr.
37, 80333 Munich, Germany}

\begin{abstract}
We propose a measure for the \char`\"{}size\char`\"{} of a Schrödinger
cat state, i.e. a quantum superposition of two many-body states with
(supposedly) macroscopically distinct properties, by counting how
many single-particle operations are needed to map one state onto the
other. This definition gives sensible results for simple, analytically
tractable cases and is consistent with a previous definition restricted
to Greenberger-Horne-Zeilinger-like states. We apply our measure to
the experimentally relevant, nontrivial example of a superconducting
three-junction flux qubit put into a superposition of left- and right-circulating
supercurrent states and find this Schrödinger cat to be surprisingly
small. 
\end{abstract}
\maketitle
\emph{Introduction}. - In his landmark 1935 paper \cite{1935_Schroedinger_Cat},
Schr\"odinger introduced the notion of entanglement, and immediately
pointed out its implications for measurement-like setups, where a
microscopic quantum superposition may be transferred into a superposition
of two {}``macroscopically distinct'' states. His metaphor of a
cat being in a superposition of {}``dead'' and {}``alive'', initially
designed just to reveal the bizarre nature of quantum mechanics, nowadays
serves as a namesake and inspiration for a whole generation of experiments
designed to test the potential limits of quantum mechanics in the
direction of the transition to the {}``macroscopic'' world, as well
as to display the experimentalists' prowess in developing applications
requiring the production of fragile superpositions involving many
particles. Recent experiments or proposals of this kind include systems
as diverse as Rydberg atoms in microwave cavities \cite{2001_07_Haroche_CavityQED},
superconducting circuits \cite{1999_08_Mooij_QubitProposal,2000_10_Mooij_FluxQubit_Experiment,2000_07_Friedman_Cat,2001_Marquardt_CooperBox,2001_BuissonHekking_Cat,2002_05_Vion_Quantronium},
optomechanical \cite{1997_04_Mancini_MirrorCat,2003_09_Marshall_QSuperposMirror}
and nanoelectromechanical \cite{2002_04_Armour_ResonatorCat} systems,
molecule interferometers \cite{1999_10_Arndt_C60_Interference}, magnetic
biomolecules \cite{1992_05_Awschalom_MQT}, and atom optical systems
\cite{2001_Julsgaard_AtomCloudEntanglement} (for a review with more
references, see Ref. \cite{2002_04_Leggett_ReviewDistance}).

\newcommand{\ket}[1]{\left|#1\right\rangle }

The obvious question of just how many particles are involved in such
a superposition has not found any general answer so far \cite{2002_04_Leggett_ReviewDistance},
and discussions of this point in relation to existing experiments
have often remained qualitative. While the number of atoms participating
in a macroscopic superposition of a $C_{60}$ molecule being at either
one of two positions separated by more than its diameter is obviously
sixty, the mere presence of a large number of particles in the system
is not sufficient in itself. This is demonstrated clearly by the example
of a single electron being shared by two atoms in a dimer, atop the
background of a large number of {}``spectator electrons'' in the
atoms' core shells. Therefore, obtaining a general measure for the
{}``size'' of a superposition of two many-body states is nontrivial,
especially for systems such as superconducting circuits, where the
relevant superimposed many-body states are not spatially separated. 

Before proposing our solution to this challenge, we state right away
that certainly more than one reasonable definition is conceivable,
depending on which features of the state are deemed important for
the superposition to be called {}``macroscopic''. Previous approaches
can be roughly grouped into two classes: Measures of the first kind
involve considering some judiciously chosen observable, evaluating
the difference between its expectation values for the two superimposed
states, and expressing the result in some appropriate {}``microscopic
units'' \cite{1980_Leggett_Disconnectivity,2002_04_Leggett_ReviewDistance}
or in units of the spread of the individual wave packets \cite{2004_06_BjoerkMana_CatSize}.
Several recent experiments have produced Schr\"odinger cats that,
by those measures, are remarkably \char`\"{}fat\char`\"{}. For example,
for the experiments in Delft\cite{2000_10_Mooij_FluxQubit_Experiment}
and SUNY\cite{2000_07_Friedman_Cat}, the clock- and counterclockwise
circulating supercurrents, whose superposition was studied, were in
the micro-Ampere range, leading to a difference of $10^{6}\mu_{B}$
or even $10^{10}\mu_{B}$ in the magnetic moments, respectively. %
\begin{figure}
\includegraphics[%
  width=1\columnwidth]{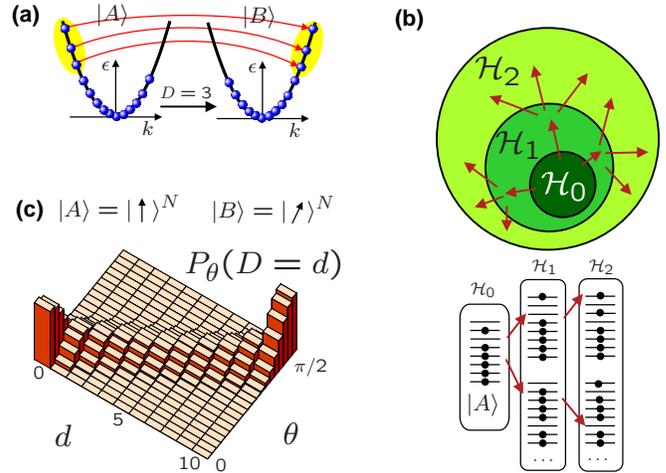}

\caption{\label{fig1}(Color online) (a) Example of normal-state persistent
currents mentioned in the text, where $D=3$ single-particle operations
are needed to turn state $\ket{A}$ into $\ket{B}$. (b) Hilbert spaces
$\mathcal{H}_{d}$ generated by repeated application of single-particle
operators. (c) Probability distribution $P_{\theta}(D=d)$ for the
distance $D$ between two BEC states or between the two components
of a generalized GHZ state, as a function of the angle between the
corresponding single-particle states, for $N=10$ particles, see Eq.
(\ref{lambdad}).}
\end{figure}

Measures of the second kind, in contrast, try to infer how many particles
are effectively involved in those superpositions, which will be the
focus of our paper. This category comprises Leggett's {}``disconnectivity''
\cite{1980_Leggett_Disconnectivity,2002_04_Leggett_ReviewDistance}
and the measure of D\"ur, Simon and Cirac \cite{2002_11_Cirac_Cat}
(DSC). The latter applies to a class of generalized Greenberger-Horne-Zeilinger
(GHZ) states, which it compares to standard GHZ states in terms of
susceptibility to decoherence and entanglement content. 

In this paper, we propose a general definition for the size of a Schr\"odinger
cat, or the effective distance between its two constituent many-body
states, that is based on asking the following question: {}``How many
single-particle operations are needed (on average) to map one of the
two states to the other?'' (see Fig. 1 a)

We will make this definition precise using the language of second
quantization and show that in simple analytically tractable cases
it agrees with reasonable expectations and with the measure of DSC
\cite{2002_11_Cirac_Cat}. After analyzing the general features of
our measure, we evaluate it numerically for a superconducting three-junction
flux qubit, whose eigenstates we find by exact numerical diagonalization.
The resulting size turns out to be surprisingly small, which we attribute
to the fact that repeated applications of single-particle operators
quickly produce a very large Hilbert space, in which the {}``target''
many-body state can be represented accurately.

\emph{Definition of the measure}. - We start with a simple example.
Consider a clean, ballistic, single-channel metallic ring capable
of supporting normal-state persistent currents of electrons. Suppose
it is put in a superposition of two Slater determinants, $\ket{A}$
and $\ket{B}$, which differ only in the number of right- and left-moving
electrons (Fig. \ref{fig1}~a). The number of particles effectively
participating in this superposition is clearly equal to the number
of particles that have to be converted from right- to left-movers,
in order to turn one of these many-body states into the other. This
is identical to the number $D$ of single-particle operations that
have to be applied to effect this change: $\ket{B}\propto\prod_{j=1}^{D}\hat{c}_{k'_{j}}^{\dagger}\hat{c}_{k_{j}}\ket{A}$.
$ $

When turning this into a general definition, we have to realize that
the {}``target'' state $\ket{B}$ might be a superposition of components
that can be created from $\ket{A}$ by applying different numbers
$d$ of single-particle operations. In that case, we will end up with
a probability distribution $P(D=d)$, defined as the weights of these
components, for the {}``distance'' $D$ between $\ket{A}$ and $\ket{B}$
to equal $d$. Furthermore, repeated application of single-particle
operations may lead to a state that could have been created by a smaller
number of such operations (such as $\ket{B}\propto\hat{c}_{k}^{\dagger}\hat{c}_{k'}\hat{c}_{k'}^{\dagger}\hat{c}_{k}\ket{A}\propto\ket{A}$
if $n_{k}=1$ and $n_{k'}=0$). This has to be taken care of by projecting
out the states that have been reached already.

The general definition (whether for fermions or bosons) is obtained
by starting from the Hilbert space $\mathcal{H}_{0}\equiv{\rm span}\{\ket{A}\}$,
and applying iteratively the following scheme, to generate spaces
$\mathcal{H}_{1},\mathcal{H}_{2},\ldots$ (Fig. \ref{fig1}~b): Given
a Hilbert space $\mathcal{H}_{d}$, apply all possible single-particle
operators $\hat{c}_{j}^{\dagger}\hat{c}_{i}$ to all of its vectors.
Consider the span of the resulting vectors and subtract the orthogonal
projection on all previous Hilbert spaces, $\mathcal{H}_{0}\oplus\mathcal{H}_{1}\oplus\ldots\oplus\mathcal{H}_{d}$,
thereby generating $\mathcal{H}_{d+1}$. This scheme is guaranteed
to produce all vectors that can be generated from $\ket{A}$ by the
time-evolution of an arbitrary (interacting, possibly time-dependent,
but particle-conserving) Hamiltonian. Thus, we can represent the {}``target''
state $\ket{B}$ (which is assumed to have the same particle number)
as a superposition

\begin{equation}
\ket{B}=\sum_{d=0}^{\infty}\lambda_{d}\ket{v_{d}}\label{Bform}\end{equation}
of orthonormalized vectors $\ket{v_{d}}\in\mathcal{H}_{d}$. The amplitudes
$\lambda_{d}$ (defined up to a phase) produce the probability distribution
for the distance $D$ from $\ket{A}$ to $\ket{B}$,

\begin{equation}
P(D=d)\equiv|\lambda_{d}|^{2}\,,\end{equation}
from which an average distance $\bar{D}$ may be obtained.

\emph{Generalized GHZ states}. - Before discussing general features,
let us consider an important example, namely a superposition of two
non-interacting pure Bose condensates, $\ket{A}$ and $\ket{B}$,
with $N$ particles being simultaneously in the single-particle state
$\ket{\alpha}$ or $\ket{\beta}$, respectively, where $\left\langle \alpha\left|\beta\right.\right\rangle =\cos\theta$.
We can write the two BEC many-body states as $\ket{A}=(N!)^{-1/2}(\hat{c}_{1}^{\dagger})^{N}\ket{{\rm Vac}}$
and $\ket{B}=(N!)^{-1/2}(\cos\theta\hat{c}_{1}^{\dagger}+\sin\theta\hat{c}_{2}^{\dagger})^{N}\ket{{\rm Vac}}$,
with $\hat{c}_{1}^{\dagger}$ creating a particle in state $\ket{\alpha}$,
and $\hat{c}_{2}^{\dagger}$ creating a particle in the state that
defines the orthogonal direction in ${\rm span}\{\ket{\alpha},\ket{\beta}\}$
(we have dropped a potentially present, but irrelevant global phase).
Expanding $\ket{B}$, we find:

\begin{equation}
\ket{B}=\frac{1}{\sqrt{N!}}\sum_{d=0}^{N}\left(\begin{array}{c}
N\\
d\end{array}\right)\left(\sin\theta\hat{c}_{2}^{\dagger}\right)^{d}\left(\cos\theta\hat{c}_{1}^{\dagger}\right)^{N-d}\ket{{\rm Vac}}.\end{equation}
Then $\ket{v_{d}}=[d!(N-d)!]^{-1/2}\left(\hat{c}_{2}^{\dagger}\right)^{d}\left(\hat{c}_{1}^{\dagger}\right)^{N-d}\ket{{\rm Vac}}$
is a normalized state that can be reached from $\ket{A}$ in exactly
$d$ applications of the single-particle operator $\hat{c}_{2}^{\dagger}\hat{c}_{1}$,
i.e. $\ket{v_{d}}\in\mathcal{H}_{d}$. Thus, $\ket{B}$ may be represented
in the form (\ref{Bform}), with coefficients

\begin{equation}
\lambda_{d}=\sqrt{\left(\begin{array}{c}
N\\
d\end{array}\right)}\sin^{d}\theta\cos^{N-d}\theta.\label{lambdad}\end{equation}
The resulting distribution $P_{\theta}(D=d)=|\lambda_{d}|^{2}$ is
binomial (Fig. \ref{fig1}~c), with probability $p=\sin^{2}\theta=1-|\left\langle \alpha\left|\beta\right.\right\rangle |^{2}$,
and thus the average distance turns out to be $\bar{D}=Np$. It will
be maximal, $\bar{D}=N$, for two orthogonal single-particle states,
as expected. This example can be transcribed into spin-language, by
considering the states $\ket{A}=\ket{\uparrow}^{N}$ and $\ket{B}=(\cos\theta\ket{\uparrow}+\sin\theta\ket{\downarrow})^{N}$.
In that case, we have to adapt our approach by defining single-spin
operators as the single-particle operations, and replace $\hat{c}_{2}^{\dagger}\hat{c}_{1}$
by $\sum_{j=1}^{N}\hat{\sigma}_{-}^{(j)}$. Straightforward algebra
shows the results for $P(D=d)$ and $\bar{D}$ to be the same. Comparing
to DSC \cite{2002_11_Cirac_Cat}, where such generalized GHZ-states
were analyzed, we find that our result agrees precisely with theirs,
for this special class of states, to which the analysis of DSC was
restricted.

\emph{General features}. - We can prove that the Hilbert spaces $\mathcal{H}_{d}$
thus constructed do not depend on the choice of the single-particle
basis used to define the operators $\hat{c}_{i}^{\dagger}\hat{c}_{j}$.
Consider an arbitrary unitary change of basis, $\hat{c}_{i}'=\sum_{j}U_{ij}\hat{c}_{j}$.
Starting from an arbitrary vector $\ket{v}$, we want to show that
${\rm span}\{\hat{c}_{i}^{\prime\dagger}\hat{c}_{j}^{\prime}\ket{v}\}={\rm span}\{\hat{c}{}_{i}^{\dagger}\hat{c}{}_{j}\ket{v}\}$
(where $i,j$ range over the basis). Indeed, any vector $\ket{w}$
from the Hilbert space on the left-hand-side can be written as $\ket{w}=\sum_{i',j',i,j}\mu_{i',j'}U_{i'i}^{*}U_{j'j}\hat{c}_{i}^{\dagger}\hat{c}_{j}\ket{v}$,
which is an element of the right-hand-side (and vice versa). As a
result, no particular basis (e.g. position) is singled out.

We can prove as well that the distance is symmetric under interchange
of $\ket{A}$ and $\ket{B}$ for an important class of states, namely
those connected by time-reversal (such as left- and right-going current
states considered below). With respect to a position basis of real-valued
wave functions, this means $\ket{A}=\ket{B}^{*}$. In that case, since
the single-particle operators can be chosen real-valued, we have $\mathcal{H}_{d}^{A\rightarrow B}=(\mathcal{H}_{d}^{B\rightarrow A})^{*}$,
making the weights $P^{A\rightarrow B}(D=d)$ and $P^{B\rightarrow A}(D=d)$
equal. The example treated above can also be expressed in this way,
by an appropriate change of basis, with $\ket{A/B}\propto(\cos(\frac{\theta}{2})\hat{c}_{1}^{\dagger}\pm i\sin(\frac{\theta}{2})\hat{c}_{2}^{\dagger})^{N}\ket{{\rm Vac}}$.
For other, non-symmetric pairs of states $\ket{A}$,$\ket{B}$, this
is not true any longer, i.e $P^{A\rightarrow B}$ can become different
from $P^{B\rightarrow A}$. An extreme example is provided by the
states $\ket{A}=\frac{1}{\sqrt{2}}(\ket{N,0}+\ket{0,N})$ and $\ket{B}=\ket{N-1,1}$,
for $N$ bosons on two islands (with $\ket{n_{1},n_{2}}$ denoting
the particle numbers). Here $P^{A\rightarrow B}(D=1)=1$, but $P^{B\rightarrow A}(D=1)<1$,
with $P^{B\rightarrow A}(D=N-1)\neq0$. In the following, we will
restrict our discussion to time-reversed pairs of states.

\begin{figure}
\includegraphics[%
  width=1\columnwidth]{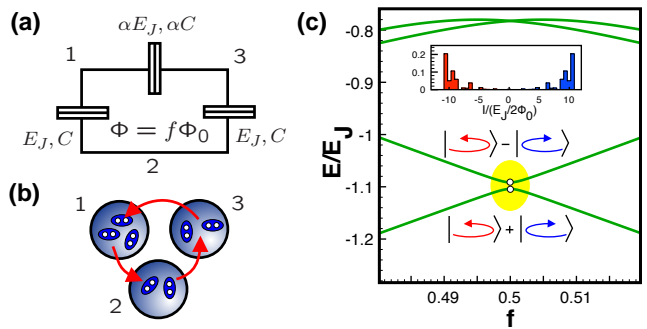}

\caption{\label{fig2}(Color online) (a) Circuit diagram of the three-junction
flux qubit. (b) Equivalent representation in the charge basis. (c)
Energy-level diagram for $E_{J}/E_{C}=20$ and $\alpha=1$, as a function
of magnetic frustration. At $f=0.5$, the ground and first excited
state are superpositions of left- and right-going current states,
$\ket{\pm I}$, the states between which we calculate the {}``distance''
$D$. The current distribution in the ground state is displayed in
the inset.}
\end{figure}

\emph{Application to superconducting circuits}. - A superconducting
circuit such as a Cooper pair box or a flux qubit device can be viewed
as a collection of metallic islands between which Cooper pairs are
allowed to tunnel coherently through Josephson junctions. Adopting
a bosonic description, we would describe tunneling by a term $\hat{c}_{i}^{\dagger}\hat{c}_{j}$,
where $\hat{c}_{j}$ annihilates a Cooper pair on island $j$. However,
as the total {}``background'' number of Cooper pairs $\bar{n}$
on any island is very large and ultimately drops out of the calculation,
the more convenient (and standard) approach is to consider instead
operators $e^{-i\hat{\varphi}_{j}}=\sum_{n}\ket{n-1}_{j}\left\langle n\right|_{j}$
that reduce the number of Cooper pairs on island $j$ by exactly one.
Then, the tunneling term becomes $\bar{n}^{-1}\hat{c}_{i}^{\dagger}\hat{c}_{j}\mapsto e^{i(\hat{\varphi}_{i}-\hat{\varphi}_{j})}$,
while the total electrostatic energy may be expressed by the number
operators $\hat{n}_{j}$ that count the number of excess Cooper pairs
on each island. These two types of operators define the single-particle
operators needed in our approach.

Let us now apply the measure defined above to a particular, experimentally
relevant case, the three-junction flux qubit that has been developed
in Delft \cite{1999_08_Mooij_QubitProposal,1999_12_Orlando_PersistentCurrentQubitPRB,2000_10_Mooij_FluxQubit_Experiment}.
Three superconducting islands are connected by tunnel junctions (Fig.
\ref{fig2}), where the tunneling amplitude is given by the Josephson
coupling $E_{J}$, and the charging energy $E_{C}=e^{2}/2C$ is determined
by the capacitance $C$ of the junctions. One of the junctions is
smaller by a factor of $\alpha$, introducing an asymmetry that is
important for the operation of the device as a qubit. The tunneling
term in the Hamiltonian is given by

\begin{equation}
\hat{H}_{J}=-\frac{E_{J}}{2}\left(e^{i(\hat{\varphi}_{2}-\hat{\varphi}_{1})}+e^{i(\hat{\varphi}_{3}-\hat{\varphi}_{2})}+\alpha e^{i(\hat{\varphi}_{1}-\hat{\varphi}_{3}+\theta)}+{\rm h.c.}\right),\end{equation}
where the externally applied magnetic flux $\Phi=f\Phi_{0}$ is expressed
in units of the flux quantum $\Phi_{0}=h/(2|e|)$ to define the frustration
$f$ that enters the extra tunneling phase $\theta=2\pi f$. The charging
Hamiltonian is

\begin{equation}
\hat{H}_{{\rm ch}}=\frac{1}{2C}\left(\hat{Q}_{1}^{2}+\hat{Q}_{3}^{2}-\frac{(\hat{Q}_{1}-\hat{Q}_{3})^{2}}{2+1/\alpha}\right),\end{equation}
with $\hat{Q}_{j}=2e\hat{n}_{j}$ and the restriction $\sum_{j=1}^{3}\hat{Q}_{j}=0$.
For simplicity, we have neglected the small effects of the self-inductance
and external gate electrodes.

At $f=0.5$, the classical left- and right-going current states are
degenerate in energy, and quantum tunneling (via the charging term)
leads to an avoided crossing, with the ground and first excited state
becoming superpositions of the two current states. We diagonalize
the current operator $\hat{I}=-\partial\hat{H}/\partial\Phi$ in the
two-dimensional subspace of the ground- and first excited states,
which results in eigenvalues $\pm I$ belonging to the two counterpropagating
current states $\ket{\pm I}$. Whenever the excited and the ground
state are well removed from higher lying levels (as should be the
case in a flux qubit), an equivalent way of finding $\ket{\pm I}$
is to write the ground state as a superposition of current operator
eigenstates in the full Hilbert space, and keeping only the contributions
with positive or negative current eigenvalues, respectively (as indicated
in the inset of Fig. \ref{fig2}). The distance $D$ between the states
$\ket{\pm I}$ then provides a measure of how {}``macroscopic''
the ground (or excited) state superposition is, in the sense of the
approach outlined above.

Our calculations have been performed in the charge basis, by truncating
the Hilbert space to $(2\Delta n+1)^{2}$ states $\ket{n_{1},n_{2},n_{3}}$
with $n_{1,2}=-\Delta n\ldots+\Delta n$ (and $n_{3}=-n_{1}-n_{2}$).
Exact numerical diagonalization of $\hat{H}=\hat{H}_{J}+\hat{H}_{{\rm ch}}$
yields the ground state and the first excited state, and, from them,
the current states $\ket{\pm I}$, as explained above. The approach
is then implemented by applying iteratively all possible single-particle
operators (represented as matrices), starting from $\ket{A}=\ket{+I}$.
The target state $\ket{B}=\ket{-I}$ is represented as a superposition
(\ref{Bform}) in the resulting Hilbert spaces $\mathcal{H}_{d}$,
which yields the weights $P(D=d)$.

\begin{figure}
\includegraphics[%
  width=1\columnwidth]{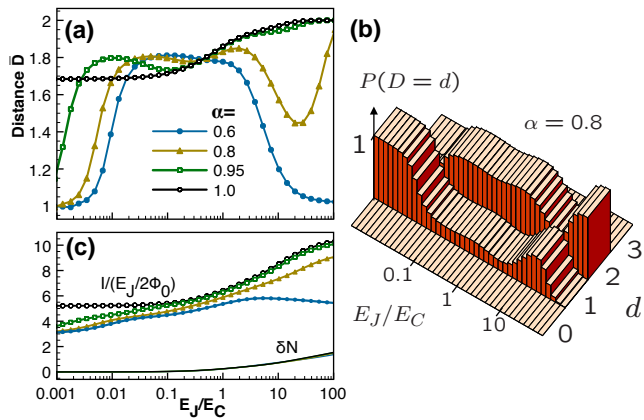}

\caption{\label{AverageDistance}(Color online) (a) Average many-body distance
$\bar{D}$ between the left- and right-going current states forming
the ground state of a three-junction flux qubit at $f=0.5$, plotted
as a function of $E_{J}/E_{C}$, for various asymmetry parameters
$\alpha$. (b) Corresponding probability distribution for $\alpha=0.8$.
(c) Magnitude $I$ of the average current in the two current states,
and average charge fluctuation $\delta N$ in the ground state (symbols
as in (a)).}
\end{figure}

Figure \ref{AverageDistance} shows the resulting average distance
$\bar{D}$ (calculated with $\Delta n=6$). The fact that $D\geq1$
is a consequence of defining the two states $\ket{A}$ and $\ket{B}$
as the eigenstates of the hermitean current operator, which makes
them orthogonal by default, thus $P(D=0)=0$. At $\alpha=1$, the
monotonic rise of $\bar{D}$ with $E_{J}/E_{C}$ is expected, as a
larger $E_{J}/E_{C}$ allows the charges on each island to fluctuate
more strongly, implying that more Cooper pairs can effectively contribute
to the current states. The nonmonotonic dependence on $E_{J}/E_{C}$
for $\alpha<1$ was unexpected, but is likely due to the fact that
smaller values of $\alpha$ tend to make the two counterpropagating
current states no longer a {}``good'' basis (the ring is broken
for $\alpha=0$). In Fig. \ref{AverageDistance} (c), we have plotted
both the expectation value of the current operator in one of the two
superimposed states, as well as the average particle number fluctuation
$\delta N$ in the ground state, where $\delta N^{2}\equiv\frac{1}{3}\sum_{j=1}^{3}\left\langle \left(\hat{n}_{j}-\left\langle \hat{n}_{j}\right\rangle \right)^{2}\right\rangle $.
Evidently, neither of these quantities can be directly correlated
to the average distance $\bar{D}$, apart from the general trend for
all of them to usually increase with increasing $E_{J}/E_{C}$. 

What is initially surprising is the fact that the distance remains
small, although the examples discussed earlier clearly show that much
larger distances may be reached in principle when applying our measure.
In contrast, the {}``disconnectivity'' for the Delft setup was estimated
\cite{2002_04_Leggett_ReviewDistance} to be on the order of $10^{6}$,
although a rigorous calculation seems very hard to do. Two reasons
underly our finding for the flux qubit: First, it appears that the
flux qubit considered here is really not that far from the Cooper
pair box. In the Cooper pair box\cite{1999_04_Nakamura_CooperPairBox},
only a single Cooper pair tunnels between two superconducting islands,
yielding $D=1$. In fact, allowing only for a small charge fluctuation
(e.g. $\Delta n=4$) on each island of the flux qubit system is sufficient
to reproduce the exact low-lying energy levels of this Hamiltonian
to high accuracy for the parameter range considered here, since the
charge fluctuations grow only slowly with $E_{J}/E_{C}$, as observed
in Fig. \ref{AverageDistance} (c) ($\delta N\sim(E_{J}/E_{C})^{1/4}$
at large $E_{J}/E_{C}$). This means from the outset that very large
values for $D$ may not be expected. Second, when analyzing the structure
of the generated Hilbert spaces $\mathcal{H}_{d}$, it becomes clear
that the dimensions of those spaces grow very fast with $d$, due
to the large number of combinations of different single particle operators
that are applied. For that reason, it turns out that the {}``target
state'' $\ket{B}=\ket{-I}$ can accurately be represented as a superposition
of vectors lying within the first few of those spaces, yielding a
rather small distance $\bar{D}$.

\emph{Outlook}. - Our measure for the {}``size'' of Schr\"odinger
cat states can be applied, in principle, to any superposition of many-body
states with constant particle number. Future challenges include the
extension to states without a fixed particle number and the comparison
to other measures, besides the DSC result \cite{2002_11_Cirac_Cat}.
In those cases where different particles couple to independent environments
(as was assumed in DSC), our measure is expected to be an indication
of the decoherence rate with which the corresponding superposition
is destroyed, and it would be interesting to verify this in specific
cases. Finally, we note that applications to many other physical systems
are in principle straightforward, though the fast growth of Hilbert
spaces may represent a significant hurdle for the direct numerical
approach used here, and more efficient approximations would be desirable. 

\emph{Acknowledgements}. - We thank I. Cirac, who drew our attention
to the question addressed here, as well as J. Clarke, K. Harmans,
J. Korsbakken, A. J. Leggett, J. Majer, B. Whaley, F. Wilhelm and
W. Zwerger for fruitful discussions. This work was supported by the
SFB 631 of the DFG.

\bibliographystyle{apsrev}
\bibliography{/Users/florian/pre/bib/shortall}

\end{document}